\documentclass[12pt]{article}

\usepackage{axodraw}
\usepackage{epsfig}
\usepackage[square,comma,sort&compress]{natbib}
\usepackage{rotating}

\def\etal{{\it et.al.}}

\begin{document}
{\flushright
\hspace{30mm} hep-ph/0410285\\
\hspace{30mm} October 2004 \\
}
\vspace{1cm}

\begin{center}
{\Large\sc {\bf Study of a Neutrino Mass Texture Generated in 
Supergravity with Bilinear R-Parity Violation}}
\vspace*{3mm}
\vspace{1cm}

{\large {Marco A. D\'\i az}$^a$, {Clemencia Mora}$^a$, 
{Alfonso R. Zerwekh}$^b$}
\vspace{1cm}

{\sl
a: Departamento de F\'\i sica, Universidad Cat\'olica de Chile,\\
Avenida Vicu\~na Mackenna 4860, Santiago, Chile \\
\vspace*{0.2cm}
b: Departamento de F\'\i sica, Universidad T\'ecnica Federico Santa 
Mar\'\i a,\\ Casilla 110-V, Valparaiso, Chile 
}
\end{center}

\vspace{2cm}

\begin{abstract}

We study a particular texture of the neutrino mass matrix generated in 
supergravity with bilinear R-Parity violation. The relatively high value 
of $\tan\beta$ makes the one-loop contribution to the neutrino mass 
matrix as important as the tree-level one. The atmospheric angle is nearly
maximal, and its deviation from maximal mixing is related to the 
smallness of the ratio between the solar and atmospheric mass scales. 
There is also a common origin for the small values of the solar and 
reactor angles, but the later is much smaller due the large mass 
ratio between the lightest two neutrinos. There is a high dependence 
of the neutrino mass differences on the scalar mass $m_0$ and the gaugino 
mass $M_{1/2}$, but a smaller one of the mixing angles on the same 
sugra parameters. Measurements of branching ratios for the neutralino 
decays can give important information on the parameters of the model. 
There are good prospects at a future Linear Collider for these 
measurements, but a more detailed analysis is necessary for the LHC.

\end{abstract}

\newpage

\section{Introduction}

With a number of experimental results in atmospheric, solar, reactor,
and accelerator neutrino physics, it has been established that 
neutrinos have mass and oscillate \cite{exp}. This is a very important 
result in its own but, in addition, it is the first direct experimental 
indication that the Standard Model (SM) needs to be modified
\cite{Grimus:2003es}.

In the SM neutrinos are massless. One popular mechanism for the generation
of neutrino masses is the see-saw mechanism, where a right handed neutrino 
field with a very large mass is added to the SM \cite{seesaw}. The 
resulting neutrino mass is inversely proportional to this large mass.
Another interesting and predictive mechanism is the radiative generation
of neutrino masses and mixing in a supersymmetric model 
\cite{Hirsch:2004he} that violates lepton number and R-Parity 
\cite{Farrar:xj} with bilinear terms in the superpotential. 
Phenomenological consequences of R-Parity violating supersymmetry are 
very distinct from R-Parity conserving models \cite{Chemtob:2004xr}.

Bilinear R-Parity breaking is an interesting mechanism for the generation 
of neutrino masses and mixing angles due to its simplicity and 
predictability \cite{Akeroyd:1997iq,Carvalho:2002bq}. It is a simple 
extension of the Minimal Supersymmetric Standard Model (MSSM) which 
includes no new fields and no new interactions. It differs from the 
MSSM in a handful of bilinear terms that violate lepton number and 
R-Parity, which cannot be eliminated with field redefinitions 
\cite{Diaz:1998vf}. Neutrino masses and mixing angles are calculable 
and agree with experimental measurements \cite{Romao:1999up,Hirsch:2000ef}.
Motivations for BRpV are for example, models with spontaneously broken 
R-Parity \cite{Nogueira:1990wz}, and a model with an anomalous horizontal
$U(1)$ symmetry \cite{Mira:2000gg}, where BRpV appears without trilinear 
R-Parity violation.

Results from SuperKamiokande \cite{exp} on atmospheric neutrinos gave
strong evidence of the $\nu_\mu-\nu_\tau$ oscillation of the 
atmospheric neutrinos with maximal or nearly maximal mixing, and
gave strong evidence against the small mixing angle solution of the 
solar neutrino problem. Results from the Sudbury Neutrino Observatory
(SNO) and the KamLAND experiment have confirmed the large mixing angle 
solution of the solar neutrino problem, showing that more than a half
of the electron-neutrinos produced at the sun oscillate into other
flavours before reaching Earth \cite{exp}. Results from the Wilkinson
Microwave Anisotropy Probe (WMAP) show temperature differences within
the microwave background radiation, which combined with results from
large scale structure give a bound on the sum of the neutrino masses
\cite{Kogut:2003et}. Finally, evidence for neutrinoless double beta
decay, if confirmed, would show the Majorana nature of neutrinos and 
the non-conservation of lepton numbers \cite{Klapdor-Kleingrothaus:2001ke}.

There are several analysis of these experimental results 
\cite{Nunokawa:2002mq}. The $3\sigma$ allowed regions for the 
neutrino parameters in \cite{Maltoni:2004ei} are
\begin{eqnarray}
1.4 \times10^{-3} < & \Delta m^2_{32} & < 
3.3 \times10^{-3}\,{\mathrm{eV}}^2
\nonumber\\
7.2 \times10^{-5} < & \Delta m^2_{21} & < 
9.1 \times10^{-5}\,{\mathrm{eV}}^2
\nonumber\\
0.52 < & \tan^2\theta_{23} & < 2.1
\label{3s5sLim}\\
0.30 < & \tan^2\theta_{12} & < 0.61
\nonumber\\
& \tan^2\theta_{13} & < 0.049
\nonumber
\end{eqnarray}
which we show for reference.

In this article, we re-analyze the possibility of having BRpV in a 
supergravity scenario, in which the scalar masses and the gaugino masses 
are universal at the GUT scale. The electroweak symmetry is broken 
radiatively but, contrary to the MSSM, sneutrinos acquire vacuum 
expectation values as well as the Higgs bosons. We give up the 
possibility that the $\epsilon_i$ and $B_i$ parameters (one for each 
lepton and analogous to the $\mu$ and $B$ term in the MSSM respectively)
are universal at the GUT scale, because otherwise there is no good 
solution for the neutrino physics compatible with experiments.

We found solutions that have not been discussed previously in the 
literature. These solutions are characterized by a large value of 
$\tan\beta$ and, therefore, the importance of one-loop contributions
to the neutrino mass matrix is enhanced.

\section{Neutrino Mass at Tree Level}

The superpotential of our BRpV model differs from the MSSM by three terms
which violate R-Parity and lepton number,
\begin{equation}
W=W_{MSSM}+\epsilon_i \hat L_i \hat H_u
\end{equation}
where $\epsilon_i$ have units of mass. We complement them with related
terms in the soft lagrangian,
\begin{equation}
{\cal L}={\cal L}_{MSSM}+B_i \epsilon_i \widetilde L_i H_u
\end{equation}
where $B_i$ also have units of mass. The presence of these terms
induce vacuum expectation values $v_i$ for the sneutrinos, which are 
calculated minimizing the scalar potential.

At tree level neutrino masses are generated via a low energy see-saw type 
mechanism. Neutrinos mix with neutralinos, and the MSSM neutralino mass 
matrix is expanded to a $7\times7$ mass matrix for the neutral fermions
\begin{equation}
{\bf M}_N=\left[\matrix{
{\cal M}_{\chi^0}& m^T \cr
m & 0 }\right]
\end{equation}
Here, ${\cal M}_{\chi^0}$ is the usual $4\times4$ neutralino mass matrix,
and $m$ is
\begin{equation}
m=\left[
\begin{array}{cccc}  
-\frac 12g^{\prime }v_1 & \frac 12gv_1 & 0 & \epsilon _1 \cr
-\frac 12g^{\prime }v_2 & \frac 12gv_2 & 0 & \epsilon _2  \cr
-\frac 12g^{\prime }v_3 & \frac 12gv_3 & 0 & \epsilon _3  \cr  
\end{array}  
\right] 
\end{equation}
which mixes the neutrinos with the neutralinos. The matrix ${\bf M}_N$
can be diagonalized by blocks, and the effective $3\times3$ neutrino mass 
matrix turns out to be equal to 
\begin{equation}
{\bf M}_{\nu}^{(0)}=-m\cdot{\cal M}_{\chi^0}^{-1}\cdot m^T=
\frac{M_1 g^2 \!+\! M_2 {g'}^2}{4\, det({\cal M}_{\chi^0})}
\left[\matrix{
\Lambda_1^2 & \Lambda_1 \Lambda_2 & \Lambda_1 \Lambda_3 \cr
 \Lambda_1 \Lambda_2 & \Lambda_2^2 & \Lambda_2 \Lambda_3 \cr
\Lambda_1 \Lambda_3 & \Lambda_2 \Lambda_3 & \Lambda_3^2
}\right]
\end{equation}
where we have defined the parameters $\Lambda_i=\mu v_i+\epsilon_i v_d$,
which are proportional to the sneutrino vacuum expectation values in the 
basis where the $\epsilon$ terms are removed from the superpotential.

This mass matrix can be diagonalized with the following two rotations.
\begin{equation}
V_{\nu}^{(0)}= 
\left(\begin{array}{ccc}
  1 &                0 &               0 \\
  0 &  \cos\theta_{23}^{(0)} & -\sin\theta_{23}^{(0)} \\
  0 &  \sin\theta_{23}^{(0)} & \cos\theta_{23}^{(0)} 
\end{array}\right) \times 
\left(\begin{array}{ccc}
  \cos\theta_{13}^{(0)} & 0 & -\sin\theta_{13}^{(0)} \\
                0 & 1 &               0 \\
  \sin\theta_{13}^{(0)} & 0 & \cos\theta_{13}^{(0)} 
\end{array}\right) ,
\end{equation}
where the reactor mixing angle in terms of the {\it alignment vector} 
${\vec \Lambda}$ is
\begin{equation}
\label{tetachooz}
\tan\theta_{13}^{(0)} = - \frac{\Lambda_1}
                   {(\Lambda_2^2+\Lambda_3^2)^{\frac{1}{2}}},
\end{equation}
and the atmospheric angle is
\begin{equation}
\label{tetatm}
\tan\theta_{23}^{(0)} = \frac{\Lambda_2}{\Lambda_3}.
\label{TreeAtmAng}
\end{equation}
As we will see later, despite the fact that tree level contribution to the 
heavy neutrino mass dominates over all loops, there are other contributions 
to the neutrino mass matrix that cannot be neglected. For this reason,
the above tree level formulas will not be enough to explain the results.

\section{Supergravity and BRpV}

In Sugra-BRpV the independent parameters are
\begin{equation}
m_0 \,, M_{1/2} \,, A_0 \,, \tan\beta \,, \mathrm{sign}(\mu) \,, 
\epsilon_i \,, \Lambda_i \,,
\label{indep}
\end{equation}
where $m_0$ is the universal scalar mass, $M_{1/2}$ is the universal 
gaugino mass, and $A_0$ is the universal trilinear coupling, valid at the 
GUT scale. In addition, $\tan\beta$ is the ratio between the Higgs vacuum
expectation values, and ${\mathrm{sign}}(\mu)$ is the sign of the 
higgsino mass parameter, both valid at the EWSB scale. Finally, 
$\epsilon_i$ are the supersymmetric BRpV parameters in the superpotential,
and $\Lambda_i$ are the parameters depending on the sneutrino vacuum 
expectation values.

We use the code SUSPECT \cite{Djouadi:2002ze} to run the two loops RGE
from the unification scale down to the weak scale. The electroweak symmetry
breaking is analogous to the MSSM, with the difference that there are 
three extra vacuum expectation values corresponding to the sneutrino vev's
$v_i$. These vev's are small and constitute a small perturbation to the
MSSM EWSB.

Despite the fact that sneutrino vev's are dependent quantities since
they are calculated from the minimization of the scalar potential, we 
remove from the group of independent parameters the $B_i$'s in favour of
$\Lambda_i=\mu v_i+\epsilon_i v_d$ as indicated in eq.~(\ref{indep}),
because they are more useful in describing the neutrino physics.

Our analysis will be centered around the SPS1 scenario in Sugra from the 
Snowmass 2001 benchmark scenarios \cite{Ghodbane:2002kg}, which is defined 
by
\begin{equation}
m_0=100\,{\mathrm{GeV}} \,,\, M_{1/2}=250\,{\mathrm{GeV}} \,, \,
A_0=-100\,{\mathrm{GeV}} \,,\, \tan\beta=10 \,,\, \mu>0
\label{sugBench}
\end{equation}
This scenario is typical of Sugra, with a neutralino LSP with a mass
$m_{\chi^0_1}=99$ GeV, and a light neutral Higgs boson with a mass just 
above the experimental limit $m_{h}=114$ GeV.

In this context we find several solutions for neutrino physics which satisfy 
the experimental constraints on the atmospheric and solar mass squared 
differences, the three mixing angles, and the mass parameter associated
with neutrino-less double beta decay \cite{Klapdor-Kleingrothaus:2003rv}. 
For illustrative purposes we single out the following
\begin{eqnarray}
\epsilon_1=-0.0004 \,,\, &\epsilon_2=0.052 \,,\, &\epsilon_3=0.051 \,,\, 
\nonumber\\
\Lambda_1=0.022 \,,\, &\Lambda_2=0.0003 \,,\, &\Lambda_3=0.039 \,,\, 
\label{epslam}
\end{eqnarray}
This solution is characterized by 
\begin{eqnarray}
\Delta m^2_{32}=2.7\times10^{-3}\,{\mathrm{eV}}^2\,, &
\Delta m^2_{21}=8.1\times10^{-5}\,{\mathrm{eV}}^2\,, &
m_{ee}=0.0036\,{\mathrm{eV}}
\nonumber\\
\tan^2\theta_{23}=0.72 \,, & \tan^2\theta_{12}=0.54 \,, & 
\tan^2\theta_{13}=0.0058
\label{nuvalues}
\end{eqnarray}
which are well inside the experimentally allowed window in 
eq.~(\ref{3s5sLim}).
We note that the random solution in eq.~(\ref{epslam}) is compatible with
$\epsilon_1=\Lambda_2=0$, {\it i.e.}, the neutrino parameters in 
eq.~(\ref{nuvalues}) are hardly changed with this replacement.

\section{Texture of Neutrino Mass Matrix}

Among the solutions to neutrino physics that we have found in our model, 
there are a few textures \cite{Smirnov:2004ju} for the effective neutrino 
mass matrix. Our study case in eq.~(\ref{epslam}) belongs to the most 
frequent one, which is
\begin{equation}
{\bf M}^{eff}_{\nu}=m\left[\matrix{
\lambda & 0 & \lambda \cr
0       & a & a       \cr
\lambda & a & 1
}\right]
\label{texture}
\end{equation}
with $a\sim 0.5-0.8$, $\lambda\sim 0.1-0.3$, and $m\sim0.02-0.04$ eV. 
To understand how this texture works we expand the neutrino masses and 
mixing angles in powers of $\lambda$. Keeping terms up to first order, 
the three neutrino masses are
\begin{eqnarray}
m_{\nu_1} &=& \lambda m+{\cal O}(\lambda^2) \cr
m_{\nu_2} &=& {\textstyle{1\over2}}m(1+a-\sqrt{5a^2-2a+1})+
{\cal O}(\lambda^2) \cr
m_{\nu_3} &=& {\textstyle{1\over2}}m(1+a+\sqrt{5a^2-2a+1})+
{\cal O}(\lambda^2)
\label{numassapp}
\end{eqnarray}
and the rotation matrix that diagonalizes the neutrino mass matrix, 
denoted $U_{PMNS}$, is
\begin{equation}
U_{PMNS}=\left[\matrix{
1 & \lambda s_{\theta}m/m_{\nu_2} & \lambda c_{\theta}m/m_{\nu_3} \cr 
\lambda/(1-a) & c_{\theta} & -s_{\theta} \cr
-\lambda/(1-a) & s_{\theta} & c_{\theta}
}\right]+{\cal O}(\lambda^2)
\end{equation}
with 
\begin{equation}
\tan 2\theta=\frac{-2a}{1-a}\,.
\label{tatmapp}
\end{equation}
In the same approximation, the atmospheric, solar, and reactor angles are 
given by
\begin{eqnarray}
\tan2\theta_{23}&=&2a/(1-a)+{\cal O}(\lambda^2) \cr
\tan\theta_{12}&=&\lambda s_{\theta}m/m_{\nu_2}+{\cal O}(\lambda^2) \cr
\sin\theta_{13}&=&\lambda c_{\theta}m/m_{\nu_3}+{\cal O}(\lambda^2)
\label{anglesapp}
\end{eqnarray}
while the atmospheric and solar mass differences are
\begin{eqnarray}
\Delta m_{32}^2&=&m^2(1+a)\sqrt{5a^2-2a+1}+{\cal O}(\lambda^2) \cr
\Delta m_{21}^2&=&{\textstyle{1\over2}}m^2\left[
1+3a^2-(1+a)\sqrt{5a^2-2a+1}\right]+{\cal O}(\lambda^2)\,.
\label{atmsolmass}
\end{eqnarray}

As an example, consider $a=1/2$ and $m=0.04$ eV. We find 
$\Delta m^2_{32}=3\sqrt{5}m^2/4\approx2.7\times10^{-3}\,{\mathrm{eV}}^2$, 
and $\Delta m^2_{21}=(7-3\sqrt{5})m^2/8\approx 5.8\times10^{-5}\,
{\mathrm{eV}}^2$, both in agreement with experiments. The third parameter 
which in this approximation does not depend on the small parameter 
$\lambda$ is the atmospheric angle, obtaining $\tan^2\theta_{23}\approx0.4$
from eq.~(\ref{tatmapp}). This value is at the lower end of the allowed 
region, nevertheless, taking $a=0.6$ we obtain 
$\tan^2\theta_{23}\approx0.5$, which is in better agreement with 
experiments.

The fact that $a$ is smaller than unity implies that the atmospheric mixing 
is not maximal. In the limit $a\rightarrow 1$, the atmospheric mixing 
approaches maximality, but the atmospheric mass 
$\Delta m^2_{atm}\rightarrow 4m^2$ which is too large if $m=0.04$ eV,
and the solar mass $\Delta m^2_{sol}\rightarrow 0$ which is too small.
Decreasing $\tan^2\theta_{atm}$ via decreasing $a$ will decrease the 
atmospheric mass scale and increase the solar one, both towards acceptable 
values. Therefore, the value of $a$ relates these three neutrino 
parameters, such that the non-maximal value for the atmospheric angle 
is connected to the smallness of the ratio between the solar mass scale 
and the atmospheric one.

The previous considerations are modified by the non zero value of 
$\lambda$. In the approximation we are working, the solar and reactor 
angles are proportional to the parameter $\lambda$, thus they are small 
quantities themselves. Nevertheless, the presence of $m_{\nu_2}$ in the 
denominator of $\tan\theta_{12}$ as opposed to $m_{\nu_3}$ in the denominator
of $\tan\theta_{13}$, makes the reactor angle much smaller than the solar 
angle. In the case $a=1/2$ and $\lambda=0.2$ we find for the solar angle 
$\tan^2\theta_{12}=0.3$, which is in the lower part of the allowed region 
and compatible with experiments. For the reactor angle we find 
$\tan^2\theta_{13}=0.017$ which is well below the experimental upper 
bound. We stress that we use the complete numerical calculation in the 
rest of the article, rather than these approximated formulas.

\section{One Loop Contributions}

All particles in the MSSM contribute to the renormalization of the 
neutralino/neutrino mass matrix. One of the most important contributions
comes from the bottom-sbottom loops. In the gauge eigenstate basis this
contribution is \cite{Diaz:2003as},
\begin{equation}
\label{eq:bsb}
\Delta\Pi_{ij}=-{{N_c m_b}\over{16\pi^2}}2s_{\tilde b}c_{\tilde b}
h_b^2\Delta B_0^{\tilde b_1\tilde b_2}\left[
{{\epsilon_i\epsilon_j}\over{\mu^2}}-\frac{a_3}{\mu} 
\left(\epsilon_i\Lambda_j+ \epsilon_j\Lambda_i\right) + 
\left( a_3^2 + \frac{ a_L a_R}{h^2_b}\right) \Lambda_i\Lambda_j \right] 
\end{equation}
where we have defined
\begin{equation}
a_R={g\over{\sqrt{2}}}\left({1\over3}t_Wa_1-a_2\right)
\,,\qquad
a_L={g\over{\sqrt{2}}}\,{2\over3}t_Wa_1
\end{equation}
\begin{equation}
a_1={{g'M_2\mu}\over{2\Delta_0}}\,,\quad
a_2=-{{gM_1\mu}\over{2\Delta_0}}\,,\quad
a_3=\frac{v_u}{4\Delta_0}(g^2M_1+g'^2M_2)\,,
\end{equation}
The main contributions to eq.~(\ref{eq:bsb}) can be understood as 
coming from the graph
\begin{center}
\vspace{-50pt} \hfill \\
\begin{picture}(200,120)(0,23) 
%
%
\ArrowLine(0,50)(40,50)
\Text(20,60)[]{$\nu_j$}
\Line(40,50)(80,50)
\Text(60,59)[]{$\widetilde H$}
\Text(88,50)[]{$h_b$}
\BCirc(40,50){3}
\Text(40,10)[]{$a_3\Lambda_j-\epsilon_j/\mu$}
\LongArrow(40,15)(40,45)
\ArrowArc(110,50)(30,180,360)
\Text(110,13)[]{$b$}
\BCirc(125,25){3}
\Line(125,22)(125,28)
\Line(122,25)(128,25)
\Text(77,70)[]{$\tilde b_R$}
\DashArrowArcn(110,50)(30,180,0){3}
\Text(110,90)[]{$\tilde b_1$}
\GCirc(95,75){3}{0}
\Text(88,84)[]{$s_{\tilde b}$}
\GCirc(125,75){3}{0}
\Text(132,84)[]{$c_{\tilde b}$}
\Text(143,70)[]{$\tilde b_L$}
\Line(140,50)(180,50)
\Text(132,50)[]{$h_b$}
\Text(160,59)[]{$\widetilde H$}
\ArrowLine(180,50)(220,50)
\Text(200,60)[]{$\nu_i$}
\BCirc(180,50){3}
\Text(180,10)[]{$a_3\Lambda_i-\epsilon_i/\mu$}
\LongArrow(180,15)(180,45)
\end{picture}
\vspace{30pt} \hfill \\
\end{center}
\vspace{10pt}
Here, neutrinos (in the gauge eigenstate basis) mix with Higgsinos who
in turn interact with the pair bottom-sbottom with a strength proportional 
to the corresponding Yukawa coupling. Full circles indicate the projection
of the sbottom mass eigenstate into right and left sbottom, which contribute
with a $\sin\theta_{\tilde b}$ and $\cos\theta_{\tilde b}$ respectively.
Open circles indicate the projection of the neutrino field onto the 
higgsino, proportional to the small parameter $a_3\Lambda_i-\epsilon_i/\mu$.
The quark propagator contributes with a factor $m_b$, and summing over 
color gives the factor $N_c$. Finally, we have in eq.~(\ref{eq:bsb})
\begin{equation}
\Delta B_0^{\tilde b_1\tilde b_2}\equiv
B_0(0,m^2_{\tilde b_1},m^2_b)-B_0(0,m^2_{\tilde b_2},m^2_b)
\end{equation}

The one-loop corrected neutrino mass matrix, in first approximation, has 
the general form
\begin{equation}
\Delta\Pi_{ij}=A\Lambda_i\Lambda_j+
B(\epsilon_i\Lambda_j+\epsilon_j\Lambda_i)+C\epsilon_i\epsilon_j
\label{deltapi}
\end{equation}
since all loop contributions can be expanded in this way. The terms of 
higher order in $\Lambda$ and $\epsilon$ have been neglected.

Considering the solutions to neutrino physics whose effective neutrino 
mass matrix has a texture of the form in eq.~(\ref{texture}), and including 
contributions from all one-loop graphs, we extract the numerical value of 
the above parameters and find 
$A\approx 7\,{\mathrm{eV/GeV}}^4$, $B\approx-0.5\,{\mathrm{eV/GeV}}^3$, and 
$C\approx 9\,{\mathrm{eV/GeV}}^2$.

Of the three parameters only $A$ gets a contribution at tree level, and 
we estimate
\begin{equation}
A^{(0)}=\frac{g^2M_1+g'^2M_2}{4\Delta_0}\approx7.6\,{\mathrm{eV/GeV}}^4
\label{Atree}
\end{equation}
Clearly, the tree level contribution to $A$ dominates over all one-loop
graphs. This is not true for $B$ and $C$ because this two parameters
are entirely generated at one-loop.

The contribution to $A$, $B$, and $C$ from the bottom-sbottom loops can be
read from eq.~(\ref{eq:bsb}). In the squark sector we have 
$m_{\tilde b_1}=492$, $m_{\tilde b_2}=538$ GeV, and 
$\sin2\theta_{\tilde b}=0.88$, which implies
\begin{equation}
C^{(1)}_{\tilde b}=-{{N_c m_b}\over{8\pi^2\mu^2}}s_{\tilde b}c_{\tilde b}
h_b^2\Delta B_0^{\tilde b_1\tilde b_2}\approx 9.8\,{\mathrm{eV/GeV}}^2
\end{equation}
This result is very close to the actual numerical value, and underlines the 
fact that the bottom-sbottom loops are very important in this particular 
scenario.

Considering that the value for $B$, in the supergravity model we are 
working with, is much smaller than $A$ and $C$, we might in first 
approximation neglect it in eq.~(\ref{deltapi}). In this case, for
the neutrino solution in eq.~(\ref{epslam}) we obtain the following 
approximated neutrino mass matrix,
\begin{equation}
{\bf M}^{eff}_{\nu}=\left[\matrix{
A\Lambda_1^2        & 0                     & A\Lambda_1\Lambda_3    \cr
0                   & C\epsilon_2^2         & C\epsilon_2\epsilon_3  \cr
A\Lambda_1\Lambda_3 & C\epsilon_2\epsilon_3 & A\Lambda_3^2+C\epsilon_3^2
}\right]
\label{Mefftheo}
\end{equation}
This form is precisely the texture observed in eq.~(\ref{texture}) 
obtained from the numerical results. Therefore, the zero in the neutrino 
mass matrix is there because $\Lambda_2,\epsilon_1\approx0$ and because 
$B$ is very small compared with $A$ and $C$. The three matrix elements 
of order $\lambda$ in eq.~(\ref{texture}) are explained by the fact 
that $\Lambda_1$ has a numerical value smaller than the other three
relevant parameters, as can be seen from eq.~(\ref{epslam}). Finally, 
the parameter $a$ in eq.~(\ref{texture}) is smaller than unity because 
$A\Lambda_3^2$ and $C\epsilon_3^2$ are comparable and of the same sign
and because $\epsilon_2\approx\epsilon_3$.

\section{Numerical Results}

In this section we study numerical results on the neutrino mass matrix, 
neutrino mass differences and mixing angles. We center our studies 
in the supergravity benchmark given in eq.~(\ref{sugBench}), although we
also explore the behavior of the neutrino parameters in the $m_0-M_{1/2}$ 
plane. We look for solutions to neutrino physics with different values
of the BRpV parameters $\epsilon_i$ and $\Lambda_i$, but concentrate our
attention in the particular solution given in eq.~(\ref{epslam}). 

\begin{figure}
\centerline{\protect\hbox{\epsfig{file=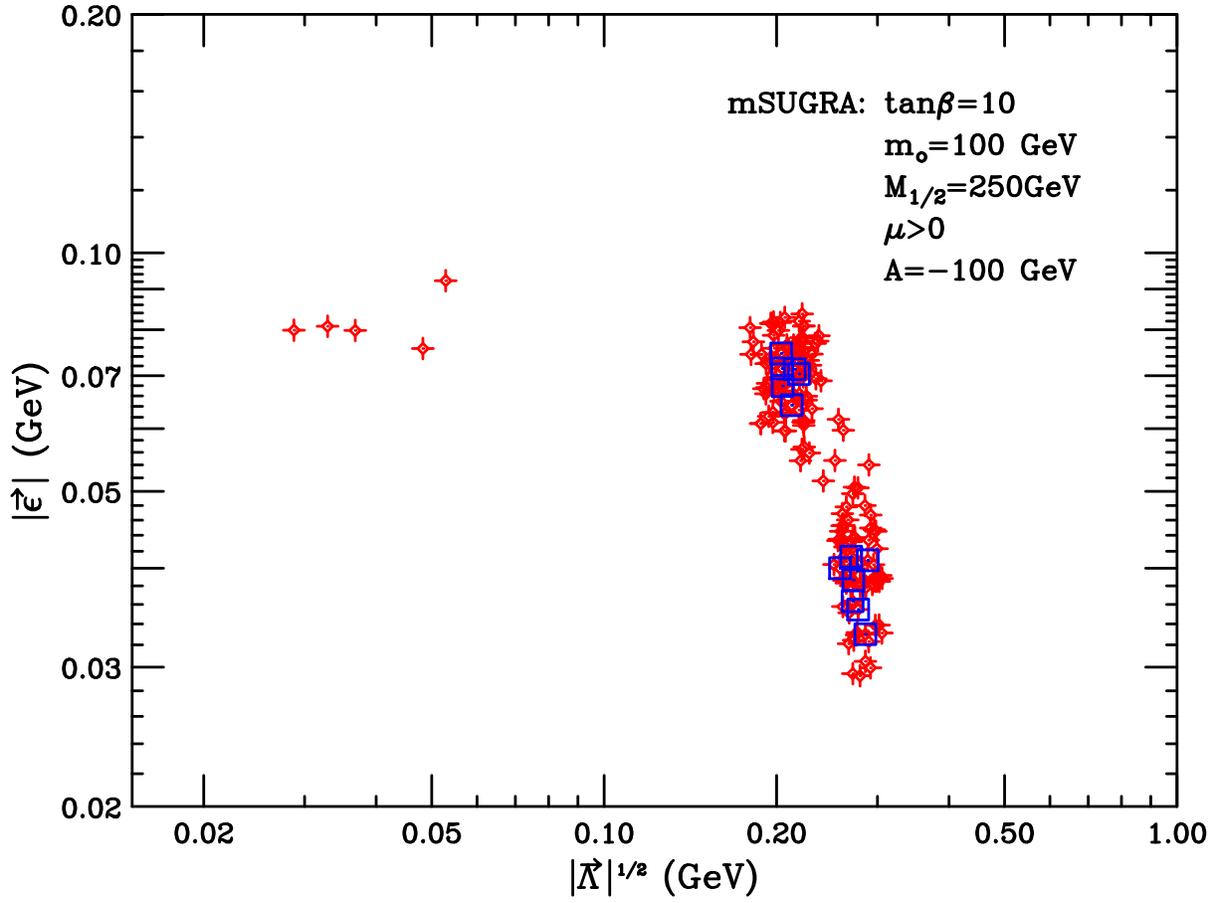,width=0.75\textwidth,angle=90}}}
\caption{\it Solutions to neutrino physics passing all experimental cuts 
described in the text, working within a particular supergravity benchmark.
}
\label{su-b1_el}
\end{figure} 
First, we consider the supergravity benchmark in eq.~(\ref{sugBench})
and randomly vary the BRpV parameters $\epsilon_i$ and $\Lambda_i$. We
look for solutions satisfying experimental restrictions on neutrino
parameters according to the $3\sigma$ intervals in eq.~(\ref{3s5sLim}), 
and also according to a relaxation of those cuts given by:
\begin{eqnarray}
1.2 \times10^{-3} < & \Delta m^2_{32} & < 
4.8 \times10^{-3}\,{\mathrm{eV}}^2
\label{relaxedcuts}\nonumber\\
0.43 < & \tan^2\theta_{23} & < 2.3
\\
5.1 \times10^{-5} < & \Delta m^2_{21} & < 
19 \times10^{-5}\,{\mathrm{eV}}^2
\nonumber
\end{eqnarray}
motivated by previous allowed regions and shown in order to compare the
effect of the improved analysis of the experimental data.

Solutions satisfying the relaxed cuts given in eq.~(\ref{relaxedcuts}) are 
displayed as green crosses in Fig.~\ref{su-b1_el}, over the plane formed by 
the absolute value of the vector $\vec\epsilon$ and the squared root of 
the absolute value of the alignment vector $\vec\Lambda$, both quantities
measured in GeV. Two distinctive regions are observed, with low and large
values of $|\vec\Lambda|$, with the low value of $|\vec\Lambda|$ solutions
harder to obtain. When the stringent cuts are implemented we find solutions
only in the region of large $|\vec\Lambda|$, and we represent them as red 
squares.

Since the tree-level neutrino mass matrix depends on $\Lambda_i$ only, and 
one-loop corrections depends on both $\Lambda_i$ and $\epsilon_i$, although 
dominated by $\epsilon_i$, the position of the solutions in the plane 
$|\vec\epsilon|$ v/s $|\vec\Lambda|$ is an indication of how important 
loop contributions are. We stress the fact, nevertheless, that increasing
values of $\tan\beta$ (which we keep constant in this study) increase
the importance of one-loop corrections, as observed in eq.~(\ref{eq:bsb})
due to the presence of the Yukawa couplings. In our case, as we will 
confirm in the following figures, the one-loop contributions to the 
neutrino mass matrix are very important.

\begin{figure}
\centerline{\protect\hbox{\epsfig{file=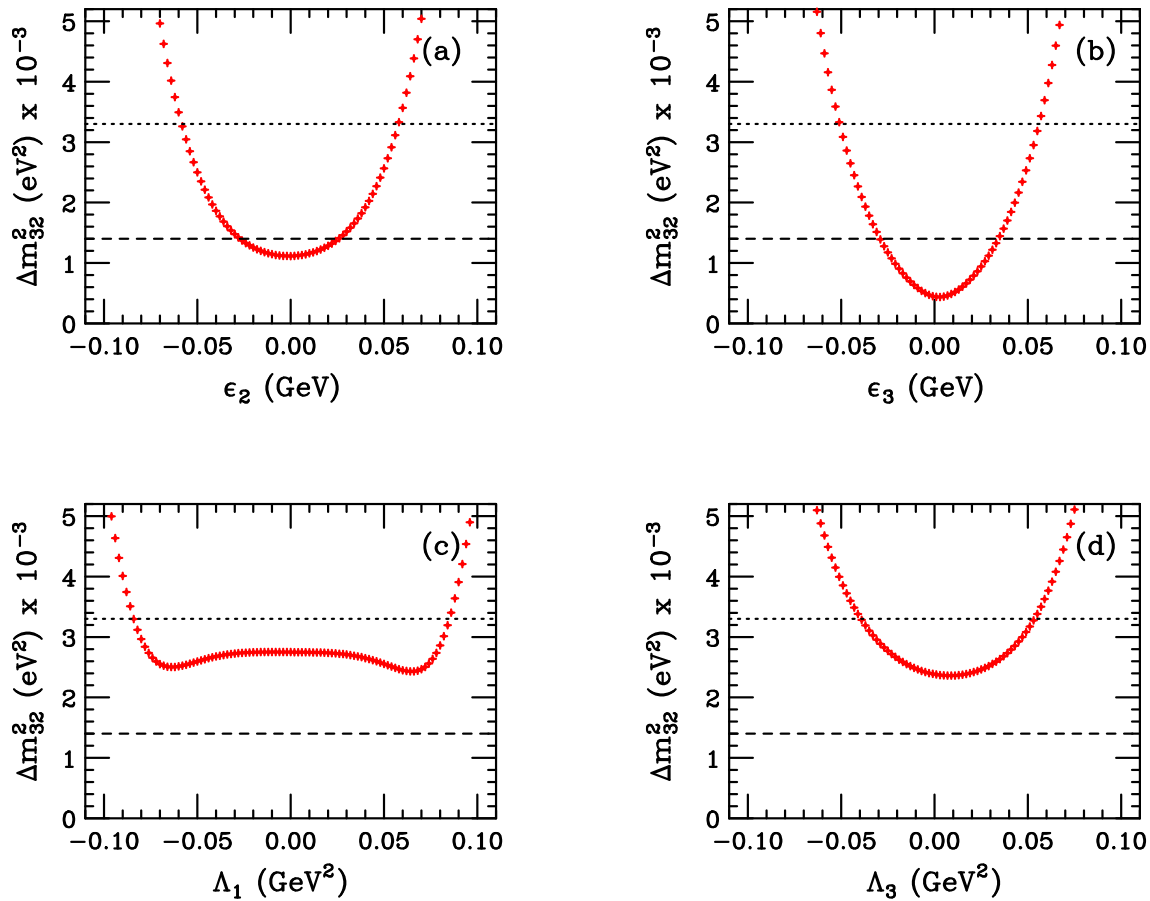,width=0.75\textwidth,angle=90}}}
\caption{\it Atmospheric mass squared difference as a function of the
four relevant BRpV parameters for the reference scenario: $\epsilon_2$,
$\epsilon_3$, $\Lambda_1$, and $\Lambda_3$.
}
\label{dmATM}
\end{figure} 
In Fig.~\ref{dmATM} we have the atmospheric mass squared difference as a 
function of the four BRpV parameters $\epsilon_2$, $\epsilon_3$,
$\Lambda_1$, and $\Lambda_3$. The neutrino mass matrix 
has the texture shown in eq.~(\ref{texture}), which implies an atmospheric
squared mass difference $\Delta m_{32}^2$ given approximately by
eq.~(\ref{atmsolmass}). 
The parameters $m$ and $a$ in eq.~(\ref{texture}) are 
$m=A\Lambda_3^2+C\epsilon_3^2$ and $a=C\epsilon_2^2/m$, as can be read 
from eq.~(\ref{Mefftheo}). The scale $m$ is quadratic in the parameters
$\Lambda_3$ and $\epsilon_3$, since the dependence of $A$ and $C$ on 
$\Lambda$'s and $\epsilon$'s is weak. The dependence of the atmospheric 
mass is obtained by replacing these expressions in eq.~(\ref{atmsolmass}), 
but when $a\approx1/2$ the atmospheric scale can be approximated even 
further obtaining, 
\begin{equation}
\Delta m_{32}^2\approx {\textstyle{\frac{3}{2}}}\sqrt{5}(A\Lambda_3^2+
C\epsilon_3^2)C\epsilon_2^2
\label{dm32app}
\end{equation}
explaining the quadratic dependency of $\Delta m_{32}^2$ on $\epsilon_2$, 
$\epsilon_3$ and $\Lambda_3$, in frames (\ref{dmATM}a), (\ref{dmATM}b), 
and (\ref{dmATM}d) respectively, and the mild dependency on $\Lambda_1$ 
(hidden in the neglected terms of order $\lambda^2$), as can be 
observed in frame (\ref{dmATM}c). The dependence on $\Lambda_1$ become
strong at high values of this parameter because in that case neglected 
terms are no longer small.

We note that using the tree level formulae in chapter 2, the atmospheric 
mass scale would be given by
$\Delta m_{32}^2\approx (A^{(0)}|\vec\Lambda|^2)^2\approx0.3\times10^{-3}$
eV${}^2$, highlighting the inadequacy of the tree level formula. On the
contrary, the approximated expression in eq.~(\ref{dm32app}) gives a
value $2.8\times10^{-3}$ eV${}^2$, which is much closer to the value
in eq.~(\ref{nuvalues}) found using the complete calculation.

\begin{figure}
\centerline{\protect\hbox{\epsfig{file=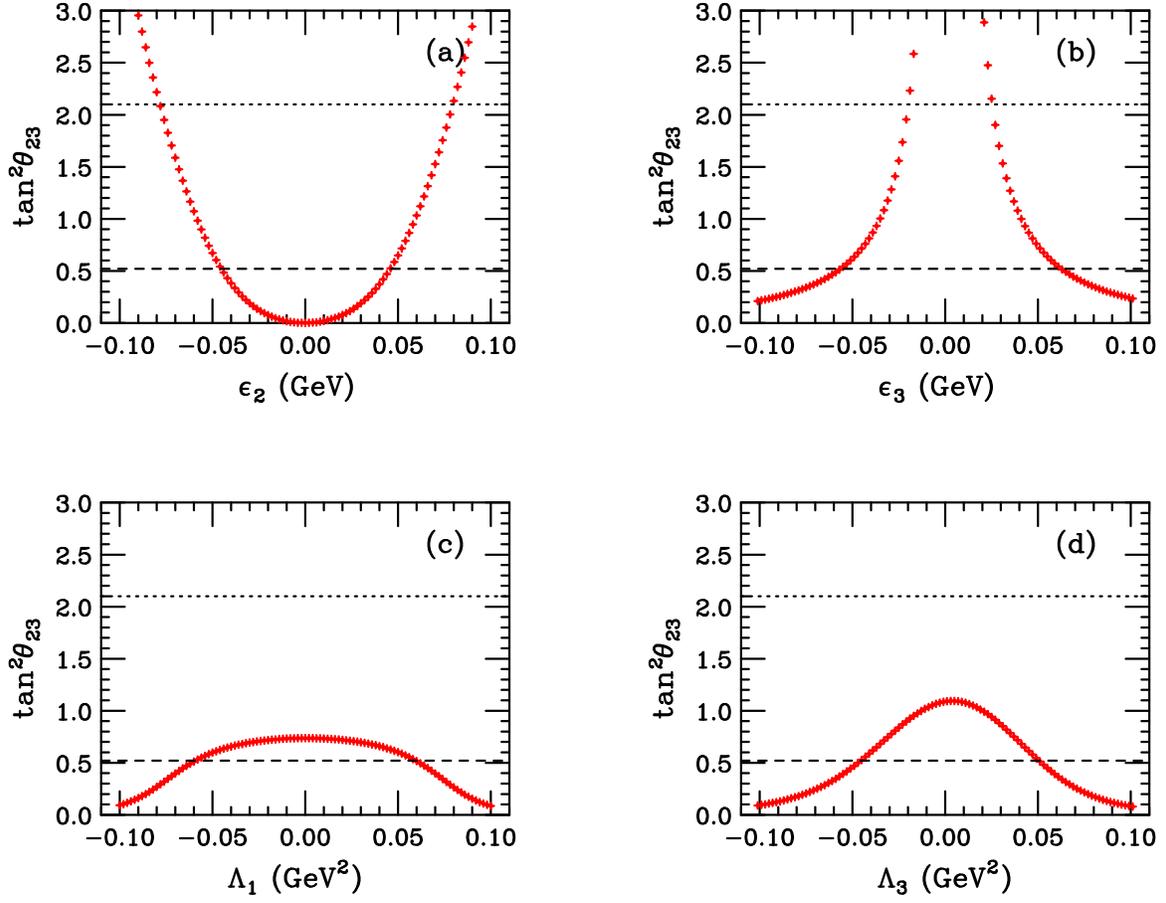,width=0.75\textwidth,angle=90}}}
\caption{\it Atmospheric angle as a function of the
four relevant BRpV parameters for the reference scenario: $\epsilon_2$,
$\epsilon_3$, $\Lambda_1$, and $\Lambda_3$.
}
\label{tanATM}
\end{figure} 
In Fig.~\ref{tanATM} we plot the tangent squared of the atmospheric angle,
$\tan^2\theta_{23}$. Using eq.~(\ref{anglesapp}), or directly from the 
mass matrix in eq.~(\ref{Mefftheo}), we find that the atmospheric angle 
satisfy
\begin{equation}
\tan2\theta_{23}\approx\frac{2C\epsilon_2\epsilon_3}
{A\Lambda_3^2+C(\epsilon_3^2-\epsilon_2^2)}
\label{AtnAngApp}
\end{equation}
This relation implies that if $\epsilon_2$ approaches zero, the 
atmospheric angle $\theta_{23}\rightarrow 0$. This behavior is confirmed 
in frame (\ref{tanATM}a). On the other hand, if $\epsilon_3$ approaches
zero then $\theta_{23}\rightarrow \pi/2$ because $C\epsilon_2^2$ is larger 
than $A\Lambda_3^2$, and this explains the divergence of $\tan\theta_{23}$ 
in frame (\ref{tanATM}b). 

In frame (\ref{tanATM}c) we see again the mild dependency of the 
atmospheric parameters on $\Lambda_2$, in this case the atmospheric angle.
If this parameter becomes very large though, neglected terms of the order
$\lambda^2$ become important. 
Finally, the dependency of the atmospheric angle on $\Lambda_3$ 
in frame (\ref{tanATM}d) can be 
understood also from eq.~(\ref{AtnAngApp}) since clearly if $|\Lambda_3|$ 
grows then $\tan\theta_{23}$ decreases.

From eq.~(\ref{TreeAtmAng}), the tree level atmospheric angle satisfy
\begin{equation}
\tan2\theta_{23}^{(0)}=\frac{2\Lambda_2\Lambda_3}{\Lambda_3^2-\Lambda_2^2}
\end{equation}
and this relation clearly misses all the influence of the one-loop graphs 
to neutrino mass matrix seen in eq.~(\ref{AtnAngApp}). Numerically, the 
approximated formula in eq.~(\ref{AtnAngApp}) gives 
$\tan^2\theta_{23}\approx0.42$, which is close to the value in
eq.~(\ref{nuvalues}). On the contrary, the tree level formula implies
$\tan^2\theta_{23}^{(0)}\approx0$.

\begin{figure}
\centerline{\protect\hbox{\epsfig{file=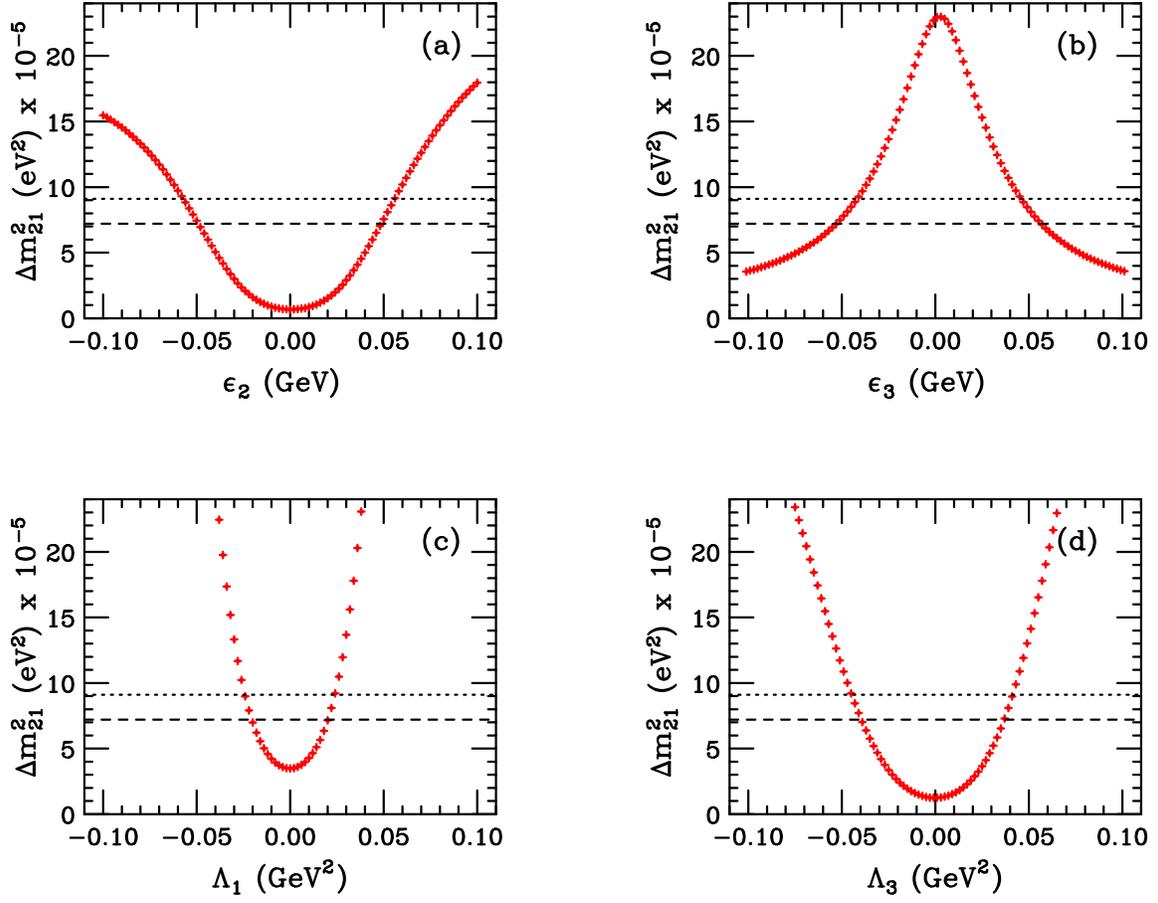,width=0.75\textwidth,angle=90}}}
\caption{\it Solar mass squared difference as a function of the
four relevant BRpV parameters for the reference scenario: $\epsilon_2$,
$\epsilon_3$, $\Lambda_1$, and $\Lambda_3$.
}
\label{dmSOL}
\end{figure} 
In Fig.~\ref{dmSOL} we plot the solar mass squared difference as a function 
of the BRpV parameters $\epsilon_2$, $\epsilon_3$, $\Lambda_1$, and 
$\Lambda_3$. In the case of $\Delta m^2_{21}$ the neglected terms of 
order $\lambda^2$ in eq.~(\ref{atmsolmass}) are numerically more important
than in the atmospheric case, therefore, predictions based on this
approximation are less accurate. 

In frame (\ref{dmSOL}a) we see the dependence of the solar mass on
$\epsilon_2$. This behavior can be understood considering that the
parameter $a$ is proportional to $\epsilon_2^2$, and when this parameter
goes to zero, the solar mass difference approaches zero like $a^2$, as seen
from eq.~(\ref{atmsolmass}).

In frame (\ref{dmSOL}b) we see how the solar mass difference depends on
$\epsilon_3$. If $\epsilon_3\rightarrow0$ then the eigenvalue 
$C\epsilon_2^2$ decouples and becomes the heaviest neutrino. Of the other
two, one neutrino is massless, and the solar mass difference becomes equal 
to the second neutrino mass squared. A growing $\epsilon_3$ will mix the
massless neutrino with the heaviest, increasing the lightest neutrino mass,
therefore, decreasing the solar mass difference, as observed in frame
(\ref{dmSOL}b).

The dependency of the solar mass on $\Lambda_1$ and $\Lambda_3$ can be 
understood only if we go beyond the simple approximation in 
eq.~(\ref{atmsolmass}). Terms of order $\lambda^2$ introduce a dependency 
on $\Lambda_1$ and $\Lambda_3$ such that $\lambda$ approaches to zero 
when these last parameters go to zero, thus explaining the behavior
shown in frame (\ref{dmSOL}c) and (\ref{dmSOL}d).

\begin{figure}
\centerline{\protect\hbox{\epsfig{file=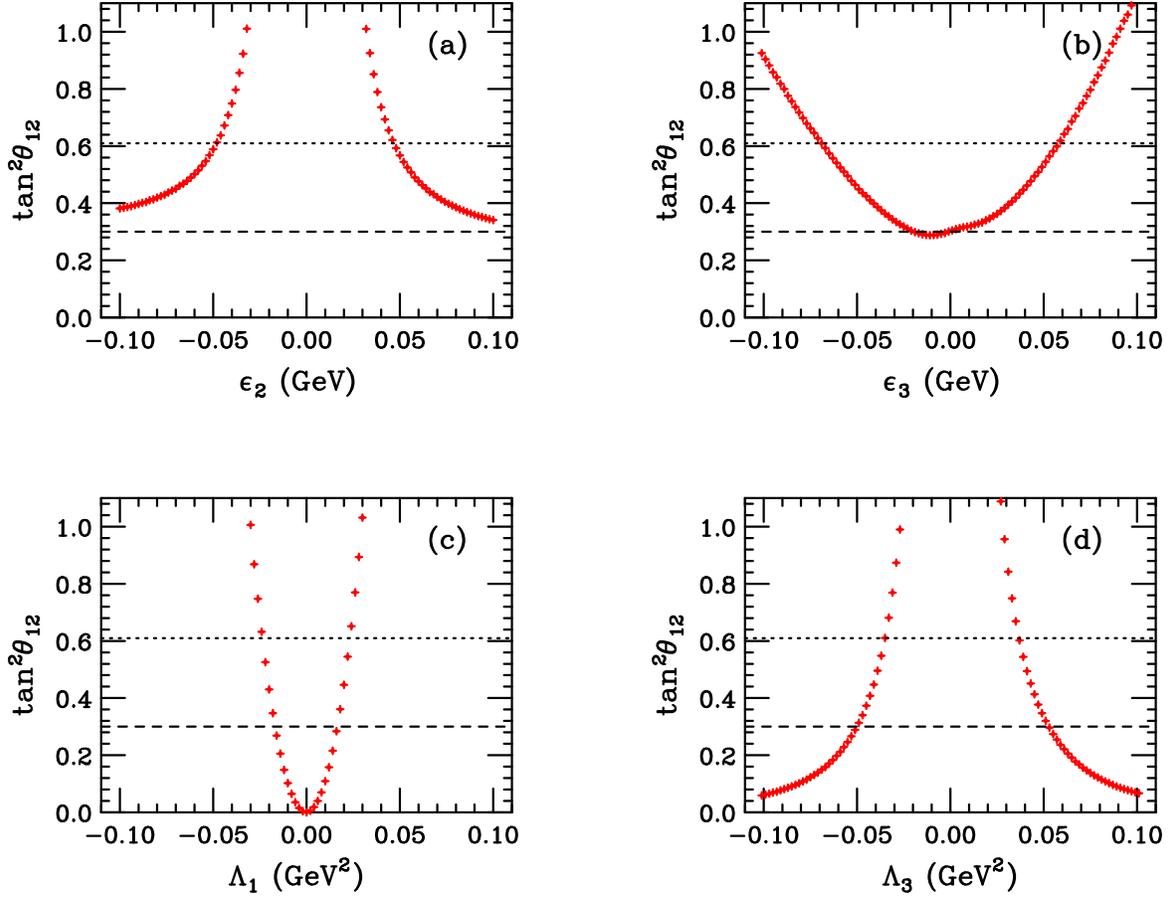,width=0.75\textwidth,angle=90}}}
\caption{\it Tangent squared of the solar angle as a function of the
four relevant BRpV parameters for the reference scenario: $\epsilon_2$,
$\epsilon_3$, $\Lambda_1$, and $\Lambda_3$.
}
\label{tanSOL}
\end{figure} 
In Fig.~\ref{tanSOL} we have the tangent squared of the solar angle,
$\tan^2\theta_{12}$, as a function of the BRpV parameters $\epsilon_2$,
$\epsilon_3$, $\Lambda_1$, and $\Lambda_3$. Working in the texture given 
in eq.~(\ref{texture}), the solar angle according to eq.~(\ref{anglesapp}) 
is approximately given by 
\begin{equation}
\tan\theta_{12}=\frac{A\Lambda_1^2}{m_{\nu_2}}\sin\theta_{23}
\label{solapp}
\end{equation}
The dependency on $\Lambda_1$ is explicit and comes from the small 
parameter $\lambda$ in eq.~(\ref{texture}). As we know from 
eqs.~(\ref{anglesapp}) and (\ref{atmsolmass}), the dependency 
of $\theta_{23}$ and $m_{\nu_2}$ on $\Lambda_1$ is weak. The behavior
of the solar angle on $\Lambda_1$ seen in frame (\ref{tanSOL}c) is thus 
understood.

The solar angle as a function of $\epsilon_2$ can also be easily 
understood noting that the parameter $a$ is directly proportional to 
$\epsilon_2^2$. According to eqs.~(\ref{numassapp}) and (\ref{anglesapp}) 
the second neutrino mass approaches zero when $a\rightarrow0$, explaining 
the divergence shown in frame (\ref{tanSOL}a). Note that $\sin\theta_{23}$ 
also approaches zero when $\epsilon_2\rightarrow0$, but slower.

The divergence of $\tan\theta_{12}$ when $\Lambda_3\rightarrow0$ is harder
to understand from the approximated expression in eq.~(\ref{solapp}), so
we go back to the effective neutrino mass matrix in eq.~(\ref{Mefftheo}).
If $\Lambda_3$ approaches zero then the upper-left element of the matrix 
decouples with a mass $A\Lambda_1^2$. On the other hand, the lower-right 
$2\times2$ sub-matrix has a zero eigenvalue, implying that 
$\theta_{12}\rightarrow\pi/2$, and therefore explaining the divergence
shown in frame (\ref{tanSOL}d).

\begin{figure}
\centerline{\protect\hbox{\epsfig{file=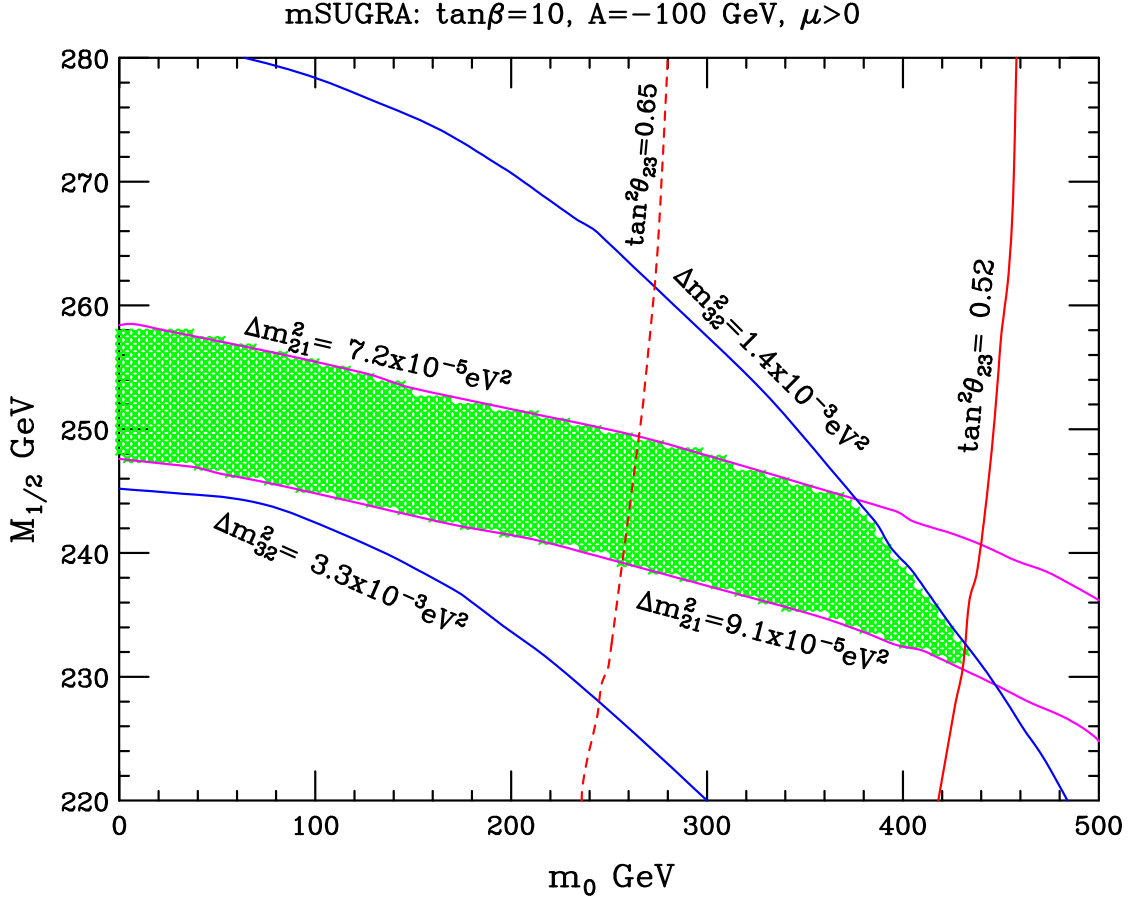,width=0.75\textwidth,angle=90}}}
\caption{\it Region of parameter space in the plane $m_0-M_{1/2}$ where
solutions to neutrino physics passing all the implemented experimental 
cuts are located. Contours of constant atmospheric mass difference and 
angle, and solar mass difference are displayed.
}
\label{m0mhalf}
\end{figure}

In Fig.~\ref{m0mhalf} we have chosen the neutrino solution given by the 
BRpV parameters in eq.~(\ref{epslam}), and vary the scalar mass $m_0$ and
the gaugino mass $M_{1/2}$, looking for solutions that satisfy all 
experimental cuts. In this case, sugra points satisfying the 
experimental restrictions on the neutrino parameters lie in the shaded
region. Solutions are concentrated in a narrow band defined by 
$M_{1/2}\approx230-260$ GeV and $m_0\approx0-400$ GeV.
We note that in BRpV the LSP need not to be the lightest neutralino,
since it is not stable anyway. For this reason, the region close to 
$m_0\approx0$ is not ruled out. 

Smaller values of $M_{1/2}$ are not possible because the atmospheric and 
solar mass differences become too large. The allowed strip is, thus, 
limited from below by the curve $\Delta m^2_{21}=9.1\times10^{-5}$ eV${}^2$. 
The dependency on $M_{1/2}$ is felt stronger by the tree level contribution 
to the parameter $A$, given in eq.~(\ref{Atree}). There we see that $A$
decreases when the gaugino mass $M_{1/2}$ increases, implying that the
atmospheric mass decreases with $M_{1/2}$, as seen in eq.~(\ref{dm32app}).
In addition, the solar mass difference is proportional to the parameter 
$m^2$, which in turn is proportional to $A$, thus, the solar mass also
decreases with the gaugino mass.

Higher values of the scalar mass $m_0$ are not allowed because the 
atmospheric angle becomes too small. The allowed strip is, therefore,
limited from the right by the contour $\tan^2\theta_{23}=0.52$. We can
understand this behavior in the following way: the parameter $C$ decreases
with increasing $m_0$ due to the Veltman's functions, and this in turn 
makes $\tan^2\theta_{23}$ to decrease with the scalar mass. High values
of the scalar mass are also limited from above because the atmospheric 
mass becomes too large. This can be explained from eq.~(\ref{dm32app})
considering that the parameter $C$ decreases with increasing $m_0$.

Higher values of $M_{1/2}$ are not possible because the solar mass becomes
too small, therefore, the allowed stripe is limited from above by the line
$\Delta m^2_{21}=7.2\times10^{-5}$ eV${}^2$. As we already mentioned, the 
solar mass difference is proportional to the parameter $m^2$, which in 
turn is proportional to $A$, and we already know that $A$ decreases with 
increasing gaugino mass $M_{1/2}$.

\begin{figure}
\centerline{\protect\hbox{\epsfig{file=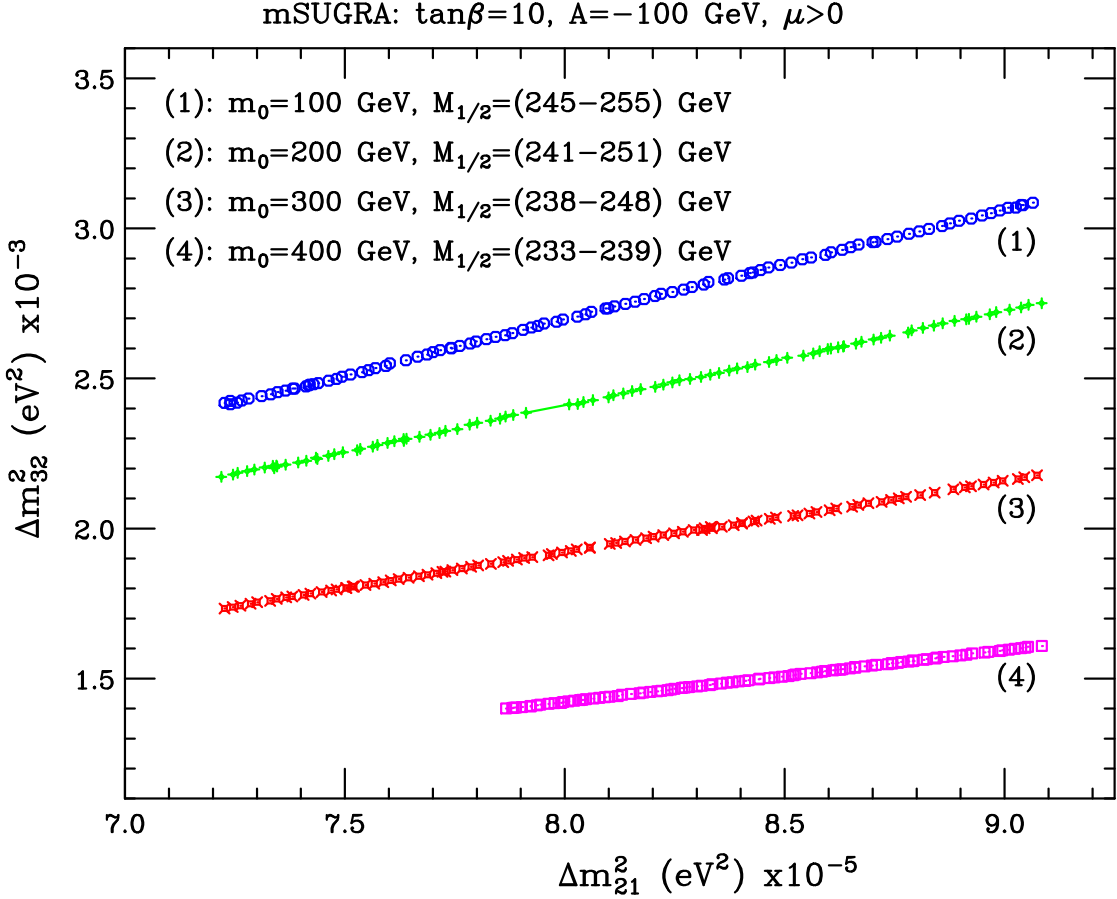,width=0.75\textwidth,angle=90}}}
\caption{\it Solutions to neutrino physics in the plane formed with the 
atmospheric and the solar mass differences. For the three different values 
of $m_0=$100, 150, and 200 GeV, we vary the gaugino mass $M_{1/2}$.
}
\label{dm32_dm21}
\end{figure}

In Fig.~\ref{dm32_dm21} we see from another point of view the dependence 
of the solar and atmospheric mass differences on the scalar mass $m_0$, 
and the gaugino mass $M_{1/2}$. In the plane formed by the atmospheric 
and solar mass differences we plot four curves defined by a constant 
value of the scalar mass $m_0=100$, 200, 300, and 400 GeV, and vary the 
gaugino mass in its allowed region, which is indicated in the figure. We 
keep fixed the values of the BRpV parameters $\epsilon_i$ and $\Lambda_i$ 
given eq.~(\ref{epslam}). The two neutrino mass differences are clearly 
proportional to each other highlighting their common origin represented 
by eq.~(\ref{deltapi}), where the parameter $A$ is controlled by tree-level
physics and the parameter $C$ is controlled by one-loop physics, and 
where both are equally important. Fig.~\ref{dm32_dm21} can be understood
further when seen in relation with Fig.~\ref{m0mhalf}.

\section{Collider Physics}

In our model, lepton number and R-Parity are not conserved. One important 
consequence is that the lightest supersymmetric particle (LSP) is not
stable, and will decay into SM particles. Since it is not stable, the LSP 
needs not to be the lightest neutralino, and whatever it is, its decays 
can be used to prove the BRpV parameters and the neutrino properties
\cite{Romao:1999up}. In the supergravity benchmark 
point considered here, the LSP is the lightest neutralino, with a mass 
$m_{\chi^0_1}=99$ GeV. 

One of the interesting decay modes of the neutralino is 
$\chi^0_1\rightarrow W^\pm l^\mp$, where $l=e, \mu, \tau$. This decay
is possible because the neutralino mixes with neutrinos which in turn
couple to the pair $W l$, and also because the charged leptons mix with
charginos and they in turn couple to the pair $\chi^0_1 W$. For this 
reason, the relevant couplings in this decay are in general very dependent
on $\epsilon_i$ and $\Lambda_i$. 

\begin{figure}
\centerline{\protect\hbox{\epsfig{file=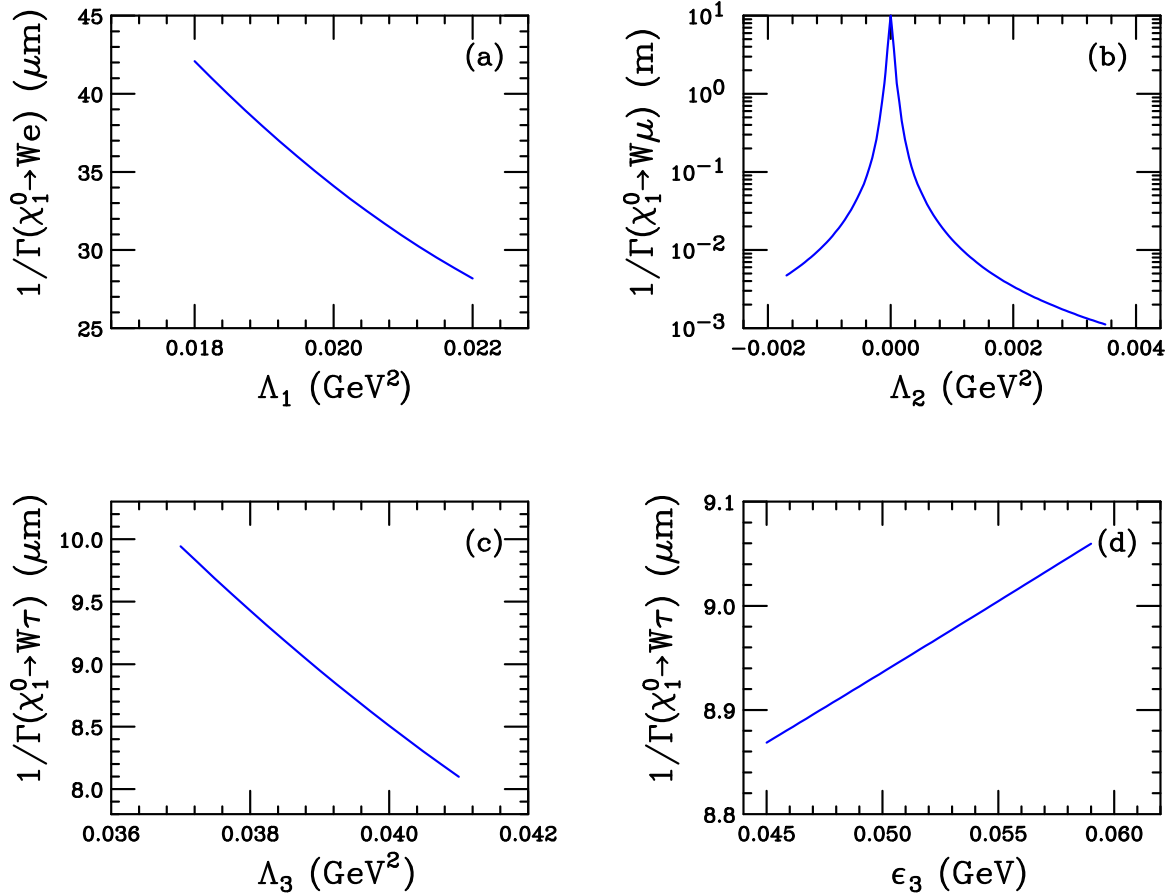,width=0.75\textwidth,angle=90}}}
\caption{\it Partial decay width of a neutralino into a $W$ and a lepton,
measured in units of distance.
}
\label{X1_g1}
\end{figure} 
In Fig.~\ref{X1_g1} we plot the inverse of the partial decay width
(multiplied by the velocity of light to convert it into a distance) as
a function of the most relevant BRpV parameters. In frame (\ref{X1_g1}a)
we see the inverse of $\Gamma(\chi^0_1\rightarrow We)$ as a function
of $\Lambda_1$. In fact, for all practical purposes, the decay rate 
into electrons depends {\it only} on $\Lambda_1$. Since in first 
approximation, the coupling is proportional to $\Lambda_1$, the inverse
of the decay rate behaves like $\Lambda_1^{-2}$, and this is seen in the 
figure. The values of $\Lambda_1$ are limited by the solar parameters.
The inverse of the partial decay rate $\chi^0_1\rightarrow We$ is of the 
order of $30-40\,\mu m$, and it's an important part of the total decay rate.

In frame (\ref{X1_g1}b) we have the inverse of 
$\Gamma(\chi^0_1\rightarrow W\mu)$ as a function of $\Lambda_2$, and 
similarly to the previous case, the decay rate into muons depends 
practically only on $\Lambda_2$. In our reference model in 
eq.~(\ref{epslam}) we have $\Lambda_2\approx0$, but values indicated
in the figure are also compatible with neutrino physics. The coupling of 
the neutralino to $W$ and muon is proportional to $\Lambda_2$, so the
inverse of the decay rate goes like $\Lambda_2^{-2}$, and that is 
observed in frame (\ref{X1_g1}b). Depending on the value of $\Lambda_2$,
the partial decay length vary from millimeters to more than a hundred 
meters in the figure. Therefore, this partial decay rate contribute little 
to the total decay rate of the neutralino.

The inverse of $\Gamma(\chi^0_1\rightarrow W\tau)$ is plotted in frames
(\ref{X1_g1}c) and (\ref{X1_g1}d) as a function of $\Lambda_3$ and
$\epsilon_3$ respectively. The dependence on $\Lambda_3$ is stronger
and similarly to the previous cases it goes like $\Lambda_3^{-2}$. The 
dependence on $\epsilon_3$ is weaker, and the inverse decay rate 
increases with this parameter. The inverse decay rate is of the order of 
$8 \mu$m, making it the most important contribution to the total decay 
rate. Neglecting any other decay mode, the total decay rate is 
near $6 \mu$m. The ratios of branching ratios for our benchmark point
in eq.~(\ref{epslam}) are given by 
\begin{equation}
\frac{B(\chi^0_1\rightarrow W\mu)}{B(\chi^0_1\rightarrow W\tau)}=
5.9\times10^{-5}\,,\qquad 
\frac{B(\chi^0_1\rightarrow W e)}{B(\chi^0_1\rightarrow W\tau)}=0.32
\end{equation}
We note that if we increase $\Lambda_2$ by a factor 4, the first ratio of
branching ratios increase to $\sim10^{-3}$ without changing the other 
ratio, while still passing all the experimental cuts. In this way, it is 
clear that by measuring the branching ratios of the neutralinos we get 
information on the parameters of the model.

The discussion above suggests that the observation of events coming
from processes like 
$pp\rightarrow \chi^0_1 \chi^0_1\rightarrow WWe\tau$ (at the LHC) or 
$e^+ e^- \rightarrow \chi^0_1\chi^0_1\rightarrow WWe\tau$ (at the NLC) 
would make possible to measure parameters relevant for neutrino physics.

\begin{figure}
\centerline{\protect\hbox{\epsfig{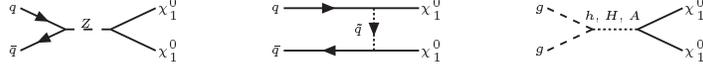}}}
\caption{\it Feynman diagrams relevant for the production of two
  neutralinos at the LHC .
}
\label{diagramas}
\end{figure} 
We use CompHEP 4.4 \cite{comphep} to calculate the production cross 
sections $\sigma(pp\rightarrow \chi^0_1 \chi^0_1)$ (LHC) and 
$\sigma(e^+ e^-\rightarrow \chi^0_1 \chi^0_1)$ (NLC at $\sqrt{s}=500$ 
GeV) at leading order. The relevant Feynman diagrams for the LHC are 
shown in Fig.~\ref{diagramas}. For the SPS1 mSugra benchmark we obtain:
\begin{eqnarray}
  \label{eq:prod2neut}
  \sigma(pp\rightarrow \chi^0_1 \chi^0_1)&=&9.8\times10^{-3} \mbox{ pb}
\nonumber\\
 \sigma(e^+ e^-\rightarrow \chi^0_1 \chi^0_1)&=&0.27 \mbox{ pb}
\end{eqnarray}
The cross sections of the whole processes were calculated multiplying
the production cross sections by the branching ratios 
$B(\chi^0_1\rightarrow W^+ e^-)$ and $B(\chi^0_1\rightarrow W^+\tau^-)$. 
Their values, for the set of parameters we have chosen, are:
\begin{eqnarray}
  \label{eq:BR}
  B(\chi^0_1\rightarrow W^+ e^-)&=&1.9\times 10^{-2}
\nonumber\\
B(\chi^0_1\rightarrow W^+\tau^-)&=&5.9\times 10^{-2}
\end{eqnarray}
The complete cross sections are:
\begin{eqnarray}
  \label{eq:completecrosssections}
  \sigma(pp\rightarrow \chi^0_1 \chi^0_1\rightarrow W^+ W^+ e^-
 \tau^-) &=&1.1\times10^{-5} \mbox{ pb}
\nonumber\\ 
 \sigma(e^+ e^-\rightarrow \chi^0_1 \chi^0_1\rightarrow W^+ W^+ e^-
 \tau^-)&=&3.0\times 10^{-4}\mbox{ pb}
\end{eqnarray}
On the other hand, the main source of background comes from the 
production of four $W$'s with two of them decaying leptonically. We 
calculated those processes using CompHEP and we found:
\begin{eqnarray}
  \label{eq:background}
  \sigma(pp\rightarrow WWWW \rightarrow W^+ W^+ e^-
 \tau^- \bar{\nu_e} \bar{\nu_{\tau}}) &=&6.5\times10^{-6} \mbox{ pb}
\nonumber\\ 
 \sigma(e^+ e^-\rightarrow WWWW \rightarrow W^+ W^+ e^-
 \tau^- \bar{\nu_e} \bar{\nu_{\tau}})&=&1.6\times 10^{-6}\mbox{ pb}
\end{eqnarray}
Assuming a luminosity of $10^5$ pb/year at, both, the LHC and the NLC
we expect 1 signal event per year at the LHC and 30 signal events per
year. Nevertheless we are not interested on the charge of the final
leptons, we only require that one lepton belongs to the first family
and the other to the third one, so the total number of signal events
are obtained by multiplying the above results by four. 

We remark that while the background is small in both cases, the number of
signal events at the LHC is also small and a more detailed analysis is
required. On the other hand the NLC appears as very auspicious
environment for studying this model.


\section{Conclusions}

We have re-examined the possibility of generating neutrino masses and
mixing angles in Supergravity with bilinear R-Parity violation. We found
solutions with a relatively large value of $\tan\beta$, such that one-loop
contributions to the neutralino mass matrix are as important as tree-level
contributions. The 
heaviest neutrino mass is still generated mainly at tree-level, but the 
other two masses and the three mixing angles are strongly affected by 
loops. In particular, the tree level approximations for the mixing angles
give completely erroneous results.

We concentrate our study on a texture for the neutrino mass matrix which
is common among our solutions, and on one particular solution corresponding 
to this texture. The atmospheric mixing is nearly maximal, and the deviation 
of the parameter $\tan^2\theta_{23}$ from unity is related to the smallness 
of the ratio between the solar and atmospheric mass scales
$\Delta m^2_{sol}/\Delta m^2_{atm}$. In addition,
the solar and reactor angles are both small because of the small parameter
$\lambda$, which in turn is small because $\Lambda_1/\Lambda_3<1$. 
Nevertheless, the reactor angle is much smaller than the solar 
angle because the second neutrino mass is much larger than the third one.

We have shown how the neutrino observables depend on the BRpV parameters
$\epsilon_i$ and $\Lambda_i$, and this dependency can be understood in 
terms of simple approximations in terms of parameters $A$, $B$, and $C$,
where all the complication of the one-loop contributions is concentrated. 
The dependency on $\epsilon_i$ and $\Lambda_i$ is strong, and it is not
clear a priori that a solution is for granted, due to the increasing 
precision of the measurements of the neutrino observables. It is shown 
also how these observables depend on the Sugra parameters, namely the 
universal scalar mass $m_0$ and the universal gaugino mass $M_{1/2}$. 
For the given by the values of $\epsilon_i$ and $\Lambda_i$, 
solutions lie in a narrow strip in the plane $m_0-M_{1/2}$, where the 
gaugino mass is strongly restricted by the solar and atmospheric mass 
scales, and the scalar mass by the atmospheric angle and mass scale.

Finally, we showed how the decay rates of the neutralino depend directly 
on some of the parameters $\epsilon_i$ and $\Lambda_i$. In fact 
$\Gamma(\chi^0_1\rightarrow W e)$ and $\Gamma(\chi^0_1\rightarrow W\mu)$
depend ${\it only}$ on $\Lambda_1$ and $\Lambda_2$ respectively, while
$\Gamma(\chi^0_1\rightarrow W\tau)$ depends on both $\Lambda_3$ and 
$\epsilon_3$. Measurements on branching ratios of the LSP can 
therefore give important information on the parameters of the model.
We estimated that a few events with $e^\pm\tau^\pm$ in the final state
can be observed at the LHC and about a hundred at the LC, indicating that 
a measurement of the decay rates is possible at the LC. A more
detailed analysis is necessary to estimate the expected precision of
these measurements.

\section*{Acknowledgements}  

This research was partly founded by CONICYT grant No.~1030948 and 
1040384.

\end{document}